\documentclass[%
 reprint,
 superscriptaddress,
 amsmath,amssymb,mathtools
 aps,
prb,
]{revtex4-1}

\usepackage{graphicx}
\usepackage{dcolumn}
\usepackage{bm}
\usepackage{color}
\usepackage{wasysym}
\usepackage{epstopdf}
\usepackage{physics}
\usepackage{eucal}
\usepackage{CJK}

\definecolor{drkgreen}{rgb}{0.0, 0.5, 0.0}

\begin{document}
\begin{CJK*}{UTF8}{gbsn} 

\preprint{APS/123-QED}

\title{Permutation cycles of hardcore Bose-Hubbard models on square and Kagome lattices}

\author{L. Shpani}
\affiliation{%
 Department of Physics, Clark University, Worcester, Massachusetts 01610, USA\\
}%

\author{F. Lingua}
\email{flingua@clarku.edu}
\affiliation{%
 Department of Physics, Clark University, Worcester, Massachusetts 01610, USA\\
}%

\author{W. Wang  (王巍)}
\affiliation{%
Max Planck Institute for the Physics of Complex Systems, N\"othnitzer Str. 38, 01187 Dresden, Germany\\
}%

\author{B. Capogrosso-Sansone}
\affiliation{%
 Department of Physics, Clark University, Worcester, Massachusetts 01610, USA\\
}%

\date{\today}

\begin{abstract}
In this paper, we study the statistics of permutation cycles of ground-state hardcore lattice bosons described by various two-dimensional Bose-Hubbard-type models on both square and Kagome lattices. We find that it is possible to differentiate quantum phases by the statistics of permutations cycles. Indeed, features in the permutation cycles statistics can be used to uniquely identify certain insulating phases, and are consistent with local resonances of occupation numbers in the ground-state expansion of the phase.
We also confirm that suitable quantities derived from the probability distribution of the length of permutation cycles can be used to detect superfluid to insulator phase transitions.
\end{abstract}

\maketitle

\section{Introduction}\label{intro}
In 1953 Feynman proposed to use permutation cycles to study statistical properties of superfluid $^{4}He$ \cite{Feynman1953}. This idea is based on the path-integral representation of the partition function~\cite{FeynmanBook} and, while the original application was in continuous systems, its validity extends to lattice systems as well. Several works looked at properties of permutation cycles to study superfluidity in continuous systems such as Bose gases \cite{Suto_1993} and dipolar bosons \cite{CintiBoninsegni2011, CintiBoninsegni2019}, $^{4}He$ \cite{Ceperley1984,Ceperley1986,Ceperley1989, Boninsegni2005}, parahydrogen \cite{Mazzacapo2008}, and lattice systems described by Bose-Hubbard-type models \cite{Boland2008,Ceperley1995}. Moreover, new Monte Carlo algorithms which use sampling of particle permutations have been developed for continuous \cite{Ceperley1986,Ceperley1989,Boninsegni2005} and lattice systems \cite{Evertz2003,Ceperley1995}.

To date, the main application of permutation cycles has been the study and detection of superfluidity, or lack thereof
(see e.g. \cite{Boninsegni2005,CintiBoninsegni2011,Boland2008}). In this work, we use this well-established tool and focus our attention on properties of permutation cycles in insulating phases of lattice bosons. We consider several Bose-Hubbard models for hardcore bosons on both square and Kagome lattices capable of stabilizing superfluid and a variety of  insulating phases. Our main result is that the statistical distribution of the length of permutation cycles in imaginary-time can be used to differentiate among different insulating phases.
We also confirm that the superfluid to insulator phase transition can be probed by considering suitable quantities related to the statistical distribution of permutation cycles.

This manuscript is organized as follows: in Section \ref{ModelMethod}, we introduce models and methods used in this work and review the concepts of permutation cycles; in Section~\ref{NumRes}, we present quantum Monte Carlo results and show that the statistics of permutation cycles of worldlines configurations can be used to differentiate among checkerboard (CB) solid, stripe (STR) solid, Z2 topologically-ordered phases at $1/3$ and $1/2$ filling, valence bond solids (VBS) at $1/3$ and $2/3$ filling~\cite{Isakov1997}, and superfluid (SF) phase. Finally, in Section~\ref{phtrans}, we confirm that suitable quantities derived from the probability distribution of the length of permutation cycles can be used to detect superfluid to insulator phase transitions. We conclude in Section~\ref{concl}.

\section{Model and Methods}\label{ModelMethod}

We consider the two-dimensional hardcore Bose-Hubbard model:
\begin{equation}
    \hat{H} = -t\sum_{\langle ij\rangle}a^\dag_i a_j + H_0\label{Hsq}
\end{equation}
where $H_0$ is the diagonal part of the hamiltonian in the Fock basis of spatial-modes, $\langle ij\rangle$ refers to sum over nearest neighboring sites, and $t$ is the hopping amplitude.
In this work, we consider square and Kagome lattices and the following $H_0$.
On the square lattice: (i) $H_0=V\sum_{\langle ij\rangle}n_i n_j$  at filling factor 1/2, that can stabilize a CB and a SF phase~\cite{Hebert2001}; and (ii) $H_0=V\sum_{ij}\frac{n_i n_j}{r_{ij}^3}$ which, among others, stabilizes a STR phase at filling factor $1/3$~\cite{Capogrosso2010} (here, $r_{ij}$ is the distance between site $i$ and $j$ and we set a cut-off at $r_{ij}=4$ lattice spacings).
On the Kagome lattice: (iii) $H_0=V\sum_{\hexagon}n_i n_j$, where the sum over $\hexagon$ refers to the sum between sites on the same hexagon of the Kagome lattice, that, at filling factor 1/2 and 1/3 can stabilize a SF and a $\mathbb{Z}_2$ topologically ordered phase~\cite{Isakov2011,Roychowdhury2015};
(iv) $H_0=V\sum_{\langle ij\rangle}n_i n_j$, which at filling $1/3$ and $2/3$ can stabilize a VBS \cite{Isakov1997}. Within this work, for all models, we consider periodic boundary conditions in space.

\begin{figure*}[ht!]
    \centering
    \includegraphics[width=1\textwidth]{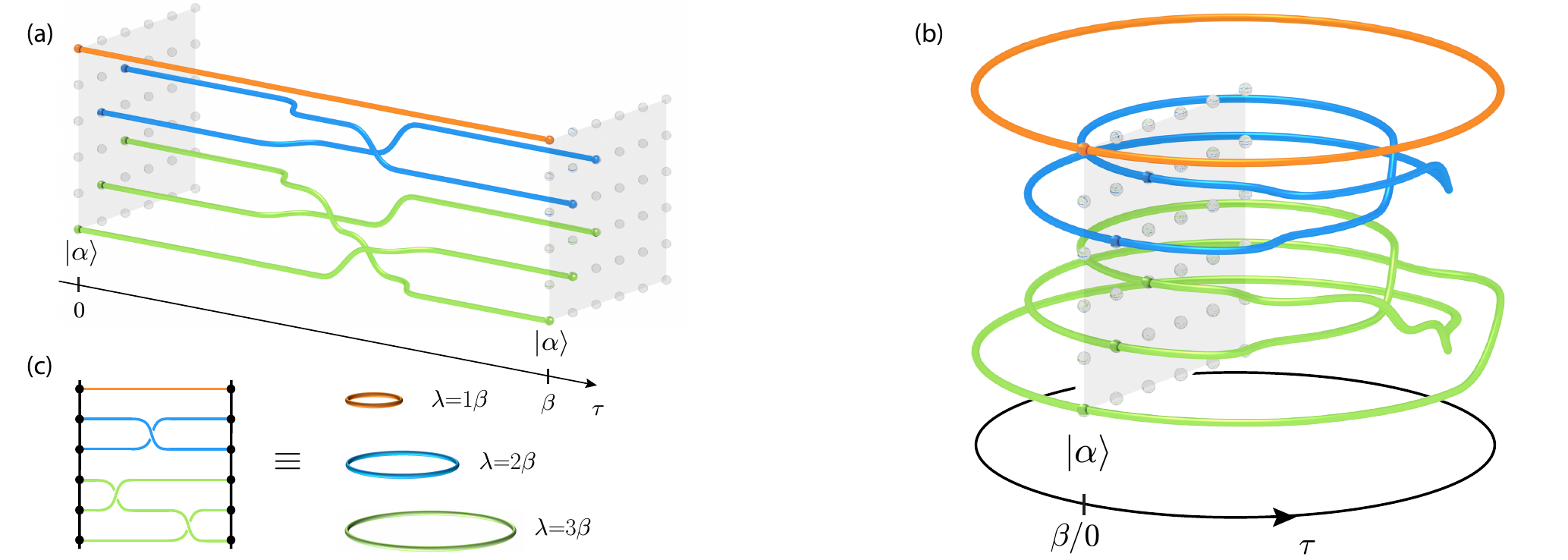}
    \caption{Example of a configuration representing the trajectories of $6$ particles in space and imaginary-time. Panel (a), the configuration is plotted in 3D, from state $|\alpha\rangle$ at $\tau=0$, to the same state $|\alpha\rangle$ at $\tau=\beta$. Panel (b), the same configuration is folded along the time direction according to periodic boundary conditions. Panel (c), the relationship between exchanges of particles and length of permutation cycles  is sketched. Particles are indistinguishable, different colors are meant to help identifying cycles  and the associated worldlines in the configuration.}
    \label{fig:2}
\end{figure*}

In the following, we present results based on path-integral quantum Monte Carlo simulations of the above models in the limit of zero temperature $k_BT=1/\beta$. We have used the worm-algorithm \cite{Prokofev1998}.
Within the path-integral formulation of the density matrix, the partition function of the system can be expressed as
\begin{equation}
\label{eq3} \mathcal{Z}_\beta= \Tr \mathrm{e}^{-\beta H}= \sum_\alpha \mel{\alpha}{\mathrm{e}^{-\beta H}}{\alpha}=\int\mathcal{D}_\chi ~\omega_\chi \; ,
\end{equation}
where $|\alpha\rangle$ are Fock states in the site occupation number representation, $\mathcal{D}_\chi$ is the measure of path-integral, and $\omega_\chi$ is the weight of a configuration $\chi$ defined as the collection of particle trajectories (worldlines) in space and imaginary-time (please refer to \cite{Ceperley1995} for a complete review on path-integral quantum Monte Carlo, and Appendix \ref{appA} for the expression of $\omega_\chi$). In Fig.~\ref{fig:2}~(a), we report a simple sketch of a configuration of 6 particles in a 6-by-6 square lattice.  Due to periodic boundary conditions in time imposed by the trace (\ref{eq3}), each Monte Carlo configuration can be folded on itself along the time direction so that one or more worldlines always form closed loops, i.e. permutation cycles.  By following worldlines in time, a permutation cycle is completed once the starting point is reached again (see Fig.~\ref{fig:2}~(b)). Permutation cycles are related to exchanges of identical particles~\cite{Feynman1953,Ceperley1995}.
We sketch these concepts in Fig.~\ref{fig:2}. The folding of the Monte Carlo configuration sketched in panel (a) is shown in panel (b), resulting in the formation of permutation cycles. In panel (c), we highlight how the exchange of particles is related to the length of permutation cycles.  Colors are only meant to help identify permutation cycles and corresponding worldlines in the configuration.
The length of a permutation cycle is always an integer multiple of $\beta$ equal to the number of worldlines participating in it. For example, a $1\beta$-long permutation cycle (orange), corresponds to a worldline glued to itself,
while, a $2\beta$-long permutation cycle  (blue), $3\beta$-long permutation cycle (green) or longer, correspond to the situation where two, three, or more worldlines are glued to each other (see panel (c) of Fig.~\ref{fig:2}). 

Here, we investigate the statistics of permutation cycles in the ground-state phases of the models described above. More precisely, we define the quantity:
\begin{equation}
    \vec{p}_\chi=(n_1, n_2, ..., n_N),\label{p_chi}
\end{equation}
where $n_l$ is the number of permutation cycles of length $\lambda=l\beta$ appearing in a Monte Carlo configuration, and $N$ is the total number of particles. We then compute the quantity
\begin{equation}
    p_\lambda:= \langle (\vec{p}_\chi)_\lambda \rangle\propto\int\mathcal{D}_\chi \;\;\frac{\omega_\chi}{\mathcal{Z}_\beta }(\vec{p}_\chi)_\lambda,\label{plambda}
\end{equation}
where $(\vec{p}_\chi)_\lambda$ is the component $\lambda$ of quantity (\ref{p_chi}). $p_\lambda$ carries the meaning of probability distribution of finding a permutation cycle of length $\lambda$.

It is known in the literature that the moments of the temporal winding-number distribution can be used to measure magnetic-susceptibility and compressibility~\cite{BookQMC_Lattice2016,Evertz2003,Hitchcock2006}.
Indeed, from distribution $p_\lambda$, we can define suitable probes capable of detecting the superfluid to insulating phase transition: 
(i) the standard deviation of the distribution~$p_\lambda$:
\begin{equation}
    \sigma_\lambda=\sqrt{\sum_{\lambda} p_\lambda (\lambda - \langle \lambda\rangle)^2}  \label{sigL}
\end{equation}
where $\langle \lambda\rangle=\sum_\lambda p_\lambda \lambda$ represents the average of permutation cycles' length;
(ii) the characteristic length $\lambda_0$ of the exponential fit of the tail of the distribution $p_\lambda$:
\begin{equation}
    p_\lambda\propto e^{-\lambda/\lambda_0}  \label{xiL}
\end{equation}

Note that expression (\ref{sigL}) is related to, though different than, the usual definition of compressibility in terms of temporal winding numbers (see e.g.~\cite{BookQMC_Lattice2016,Evertz2003,Hitchcock2006}).

\section{Numerical results}\label{NumRes}

\begin{figure}
    \centering
    \includegraphics[width=\columnwidth]{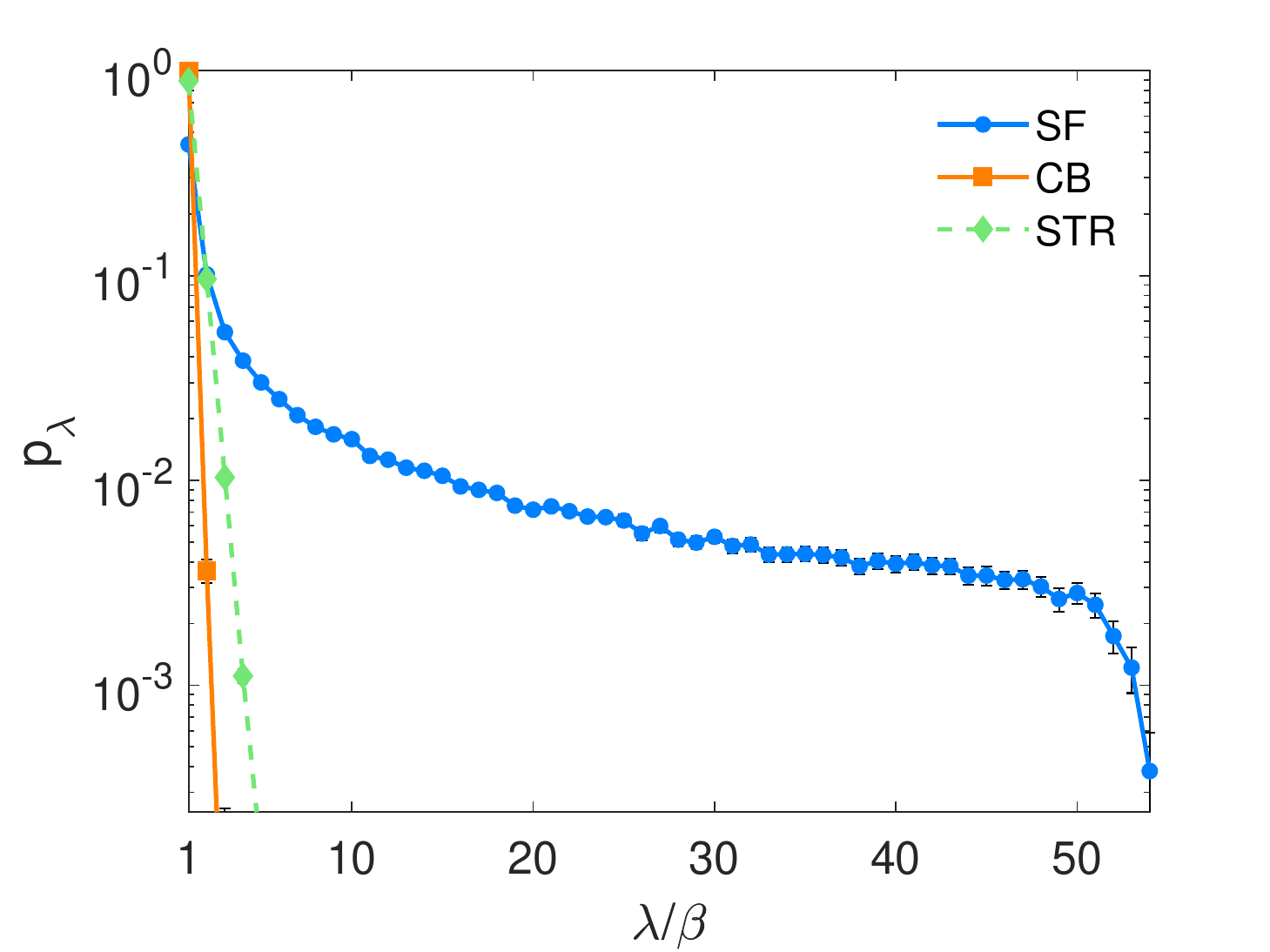}
    \caption{ Distribution $p_\lambda$ for phases stabilized on a square lattice: superfluid (blue), checkerboard (orange), stripe solid (green). Data displayed is the result of Monte Carlo simulations on a $L=10$ square lattice at $\beta/t=18$ for superfluid ($V/t=0.5$), checkerboard solid ($V/t=20$), and stripe solid ($V/t=20$). When not visible, errorbars are within symbol size.} 
    \label{fig:sqpL}
\end{figure}

\begin{figure}
    \centering
    \includegraphics[width=\columnwidth]{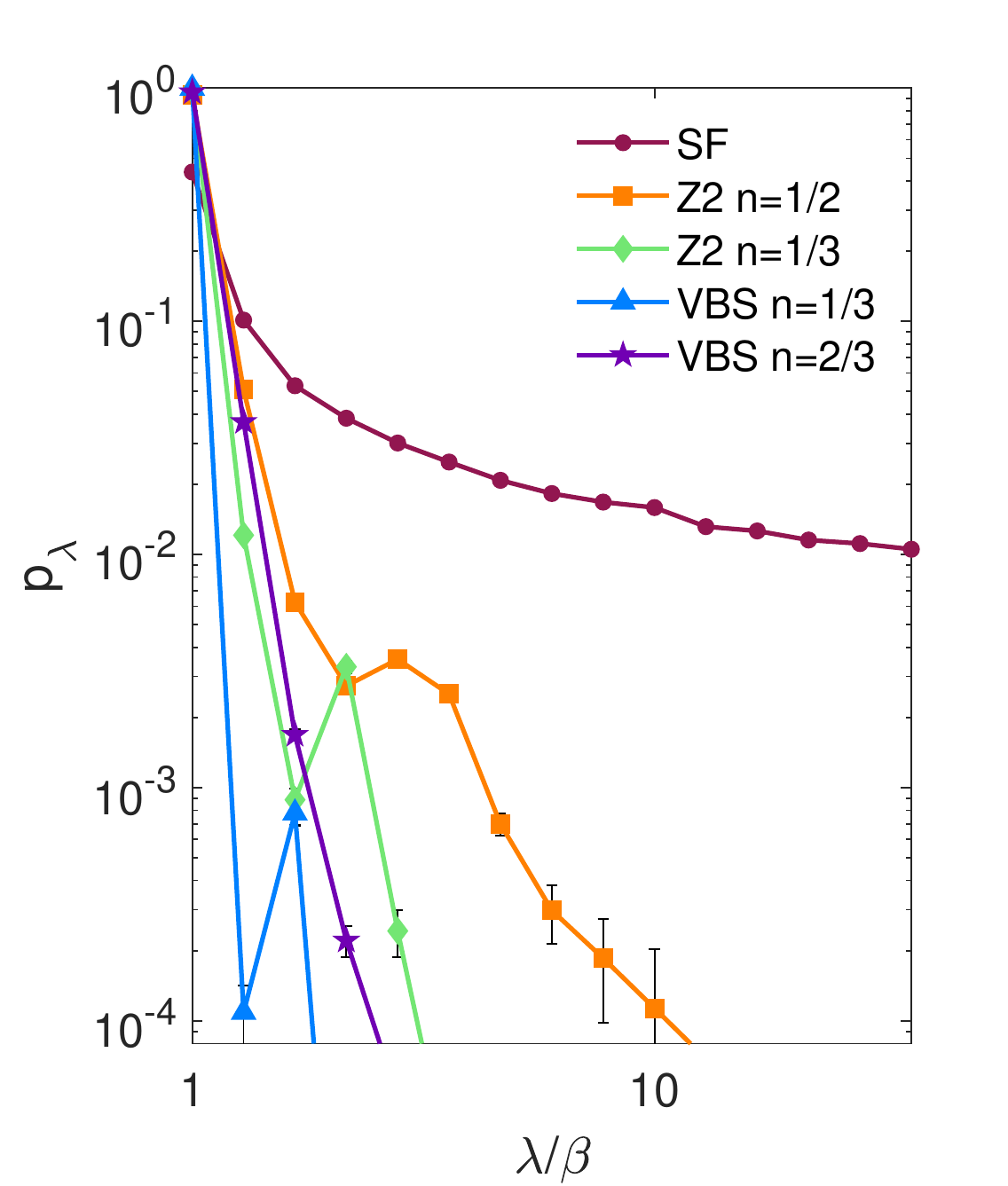}
    \caption{ Distribution $p_\lambda$ for phases stabilized on a Kagome lattice: superfluid (purple), $\mathbb{Z}_2$ topologically ordered insulator at filling $n=1/2$ (orange), and $n=1/3$ (green), valence bond solid at filling $n=1/3$ (blue), and $n=2/3$ (violet). Data displayed is the result of Monte Carlo simulations on a  $L=6$ unit cells of Kagome lattice at $\beta/t=18$ for superfluid ($V/t=0.5$), valence bond solid, and $\mathbb{Z}_2$ phases ($V/t=15$). When not visible, errorbars are within symbol size.} 
    \label{fig:kagpL}
\end{figure}

We perform Monte Carlo simulations in the insulating and superfluid phases and measure  $\vec{p}_\chi$ for a number of configurations of the order of $10^4 $ to $10^5$. From these measurements, we extract distribution $p_\lambda$. As one can see in Fig.~\ref{fig:sqpL} and Fig.~\ref{fig:kagpL}, $p_\lambda$ can be used to differentiate ground-states of phases stabilized in the different models considered.

In Fig.~\ref{fig:sqpL}, we report the distribution $p_\lambda$ of three quantum phases stabilized on the square lattice: SF phase (blue), CB solid (orange), and STR solid (green). In the SF phases, independently on $V/t$, permutation cycles of all lengths (limited by the total number of particles--the longest cycle is $N\beta$-long) have a finite probability of appearing. This is not surprising as
we expect configurations in the SF phase to be characterized by long permutation cycles~\cite{Boland2008,CintiBoninsegni2011}. Notice that this holds for both SF phases, on square and Kagome lattices (see left-most panels of Fig.~\ref{fig:pL_distr} for comparison between the two).
It is worth of notice, that, due to the upper bound on the length of permutation cycles imposed by the total number of particles (which diverges in the thermodynamic limit), $p_\lambda$ in the SF is dependent on the system-size. We have observed that, in the SF phase, the tail of the distribution $p_\lambda$ becomes flatter as the system-size increases (see appendix \ref{appB} for a discussion on the finite-size effects of the permutation cycle distribution in the SF phase). This finite-size effect is not observed in the insulating phases, for which the probability for longer permutation cycles to appear in configurations is negligible.
Qualitatively, we observe no difference between the CB and STR solids. Both distributions clearly decay exponentially (in the cycle length) to zero and have no features.

Distribution $p_\lambda$ for five quantum phases stabilized on the Kagome lattice are plotted in Fig.~\ref{fig:kagpL}: superfluid (purple), $\mathbb{Z}_2$ topologically ordered insulator at filling $n=1/2$ (orange) and $n=1/3$ (green), valence bond solid at filling $n=1/3$ (blue) and $n=2/3$ (violet). As expected, all the insulating phases exhibit negligible probability for longer permutation cycles to appear in the configuration. We note that distribution $p_\lambda$ for insulating phases on the Kagome lattice is characterized by a secondary peak whose position is different for different ground-states: $\lambda/\beta=5$ for $\mathbb{Z}_2$ at $n=1/2$, $\lambda/\beta=4$ for $\mathbb{Z}_2$ at $n=1/3$, and $\lambda/\beta=3$ for VBS at $n=1/3$.
These features are consistent with local resonances of occupation numbers in the ground-state expansion of these phases~\cite{Isakov1997,Isakov2011}.
We observe no secondary-peak for VBS at $n=2/3$, however, due to hole-particle symmetry between $n=1/3$ and $n=2/3$, we expect hole permutation cycle distribution to feature the peak at filling $n=2/3$.

\subsection{Heuristic explanation of the peaks}
The interplay between interaction, hopping amplitude, and filling factor seems to play a prominent role in determining the probability of permutation cycles. By considering a first-order perturbation theory approach within a truncated Hilbert space of lowest and first excited states in interaction energy, we provide a heuristic explanation of the peaks in $p_\lambda$ distribution. We will address separately each of the phases for which a peak appears.

\begin{figure}
    \centering
    \includegraphics[width=1\columnwidth]{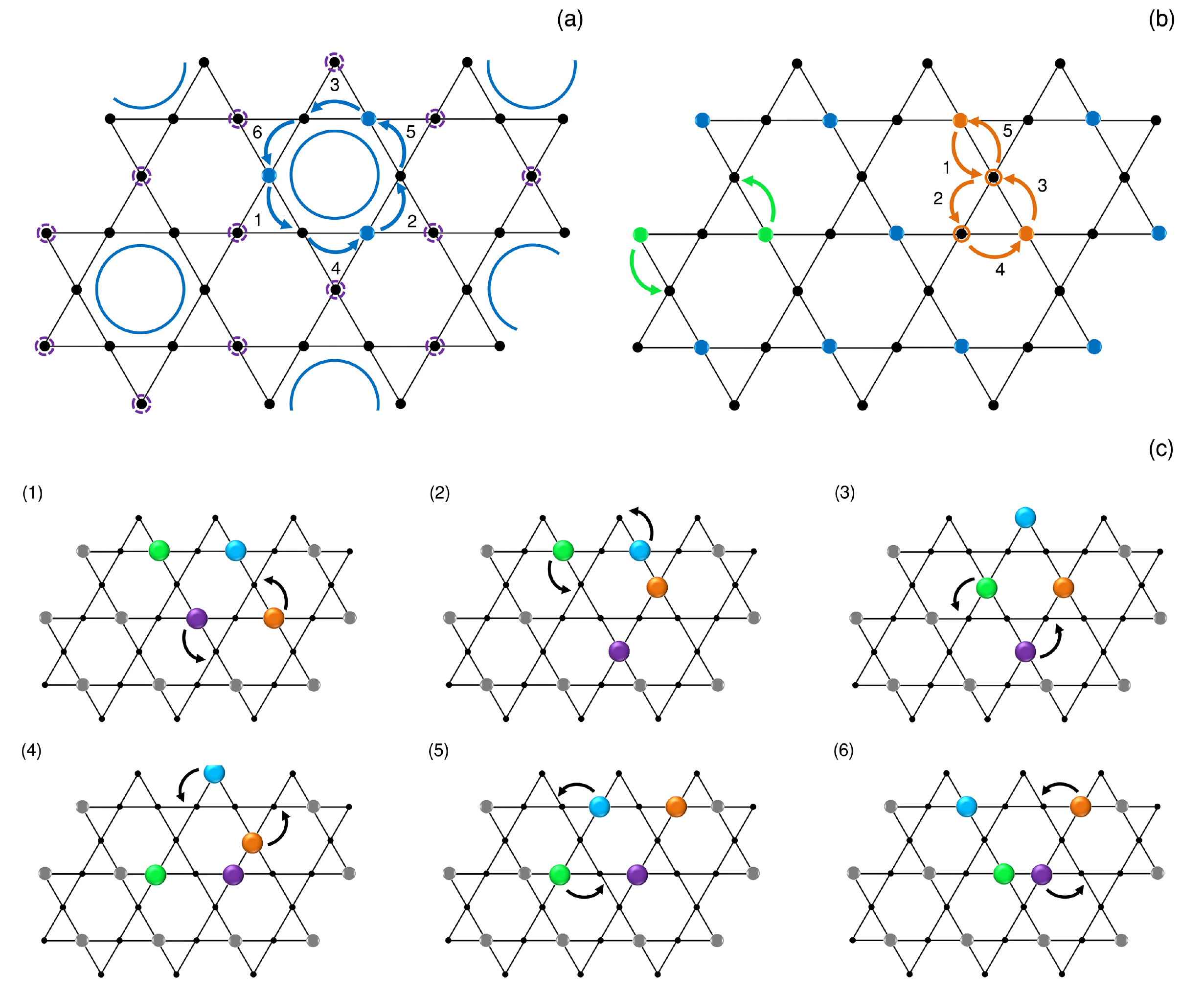}
    \caption{Panel (a): Valence Bond Solid, dashed purple circles represent the back-bone solid of holes, while the larger blue circles the resonances of three bosons. A typical hopping process leading to a permutation cycle of $\lambda/\beta=3$ is shown by the blue arrows. Panel (b): a typical minimal interaction energy configuration of $\mathcal{Z}_2$ at $1/3$ filling. Green arrows represent the zero-energy move within the effective model, while the orange arrows indicate the simplest hopping sequence for a 2-particles exchange. Panel (c): a hopping sequence contributing to the loop at $\lambda/\beta=4$ in the $\mathcal{Z}_2$ at $n=1/3$. }
    \label{fig:PCm}
\end{figure}

\emph{VBS}. In the VBS at $n=1/3$, local resonances of three bosons~\cite{Isakov1997} inside a hexagon of the Kagome lattice are protected by a solid back-bone structure of holes around the hexagon itself (see panel (a) of Fig.~\ref{fig:PCm}, where each circle represents the resonant process of the three bosons in the hexagon). In this case, it is easy to see that, within first order perturbation theory, a sequence of hopping events of the three bosons within the hexagon (see arrows in panel (a)) are responsible for a cyclic permutation of the three particles. The solid back-bone structure of holes makes exchanges of two particles (PC at $\lambda/\beta=2$) and four particles (PC at $\lambda/\beta=4$) less energetically favorable, i.e. they are higher order processes, than the three-body resonances just described.

 \emph{$Z_2$ at $n=1/3$}: A sequence of hopping events within first-order perturbation theory capable of exchanging two particles can be constructed inside a triangle of the Kagome lattice as sketched in orange in panel (b) of Fig.~\ref{fig:PCm}. Compositions of this two-particle processes can lead to exchanges of three or more particles (i.e. longer PC). From a path-integral representation point of view, the weight of such simple exchanges (this weight contributes to the calculation of the weight of a path-integral quantum Monte Carlo configuration which contains these moves) scales as (see Appendix \ref{appA}):
\begin{equation}
    \omega_\chi\propto (t\, e^{-Vd\tau})^l    \label{weight}
\end{equation}
where $l=4\cdot(m-1)$ is the number of first-excited states, at energy $\Delta E=+V$ above the minimum interaction energy configuration, visited during the exchange, and $m$ the number of particles exchanged. For example, one has $l=4$ for the simplest exchange between two particles, $l=8$ for three, and so on. Other possible low-energy moves are the ones described by the effective-model~\cite{Balents2002} on the ``bow-ties'' of the Kagome lattice (see green arrows in panel (b) of Fig.~\ref{fig:PCm}). Within the original model each bow-tie move visits a first excited state of the interaction energy before returning to a configuration of particles corresponding to a minimum of inetraction energy. At filling $n=1/3$, a well-determined sequence of this move allows a cyclic permutation of four particles around a hexagon of the Kagome lattice (see panel (c) in Fig.~\ref{fig:PCm}). This move involves $l=6$ excited states (on the other hand, using this type of moves, $m=2,3,5,6$ particle-exchanges require to visit a much larger number $l$ of excited states). It is therefore clear that the probability for $m=4$ is smaller than for $m=2$ but larger than for $m=3$ and $m=5$. This leads to a peak in the permutation cycle distribution at $\lambda/\beta=4$.

\emph{$Z_2$ at $n=1/2$}.  In this case, it is easy to see that exchanges of two particles within a triangle are the less costly, hence the most probable, as they only involve $l=2$ first-order higher-energy states in interaction energy. It is also easy to see that PC of length $\lambda/\beta=3,4,5\dots$ can be generated by subsequently exchange pair of particles with a monotonic (with the PC length) increase of visited excited states. On the other hand, we have observed that 5-particle exchanges can also be achieved by applying a sequence of bow-tie moves (as described by the effective model). These exchanges are always accompanied by 2- and/or 3-particle exchanges. This is not the case for 4- and 6-particle exchnages which instead require a much larger number of bow-tie moves, hence a much larger number of higher interaction energy states visited. We can therefore speculate that this may contribute to the peak at $\lambda/\beta=5$, i.e. more possibilities to realize the 5-particle exchanges are available compared to 4-and 6-particle exchanges. Keeping track of all possibilities with $n=1/2$ is not simple task as the number of particles in the system is larger than at filling $n=1/3$. This is beyond the scope of this work.


\subsection{Studying superfluid to insulator phase transitions}\label{phtrans}
Here, we show that the standard deviation $\sigma_\lambda$ of the probability distribution $p_\lambda$ and the characteristic length $\lambda_0$, as defined in Section \ref{ModelMethod}, capture the insulating to SF phase transition.

\begin{figure*}
    \centering
    \includegraphics[width=1\textwidth]{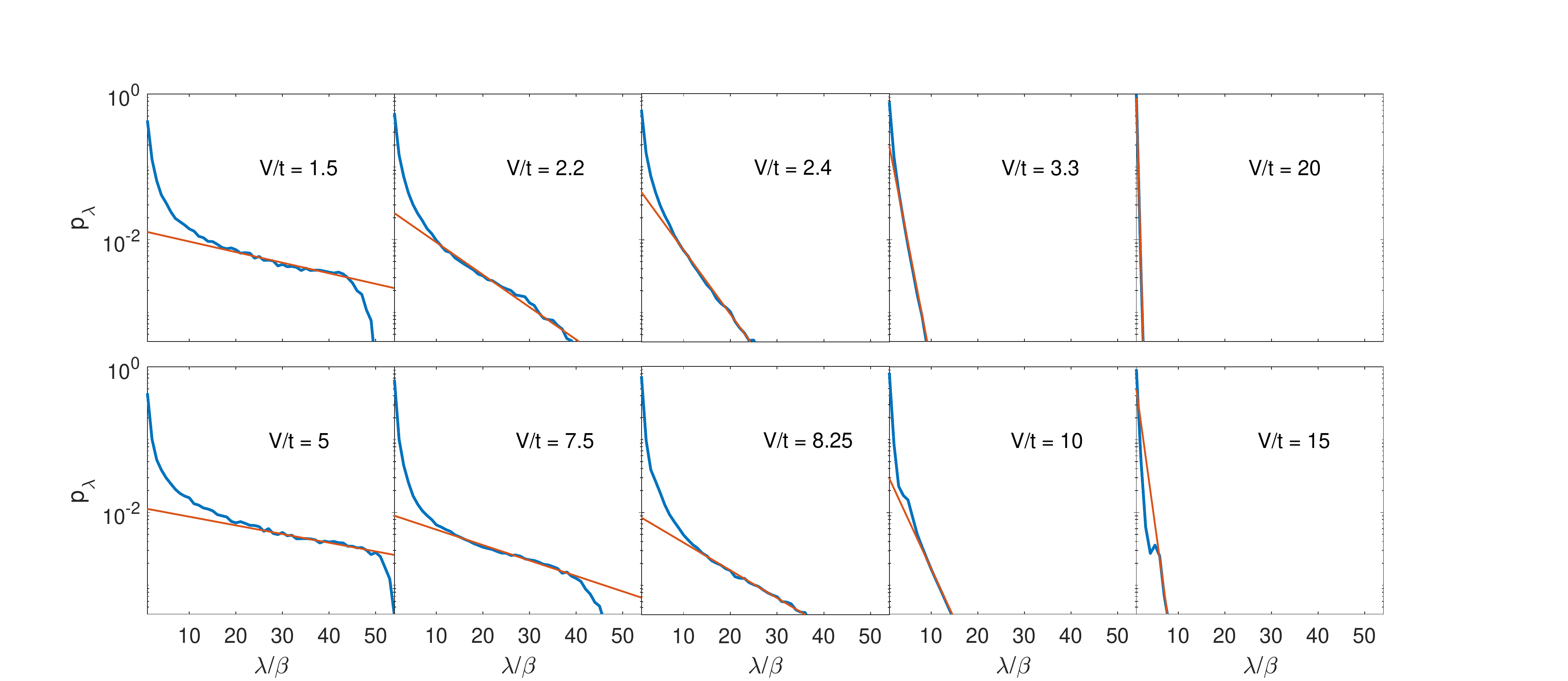}
    \caption{Distribution $p_\lambda$ (blue) and exponential fit (orange) for different values of $V/t$. Upper-panel: $p_\lambda$ across the superfluid to checkerboard transition ($(V/t)_c=2$) for system size $L=10$, $\beta=18$, and $N=50$ particles. Lower-panel: $p_\lambda$ across the superfluid to $\mathbb {Z}_2$ topologically ordered insulator transition ($(V/t)_c=7.0665$) for a system of $L=6$ unit cells, $\beta/t=18$, and $N=54$ particles.}
    \label{fig:pL_distr}
\end{figure*}

\begin{figure}
    \centering
    \includegraphics[width=1\columnwidth]{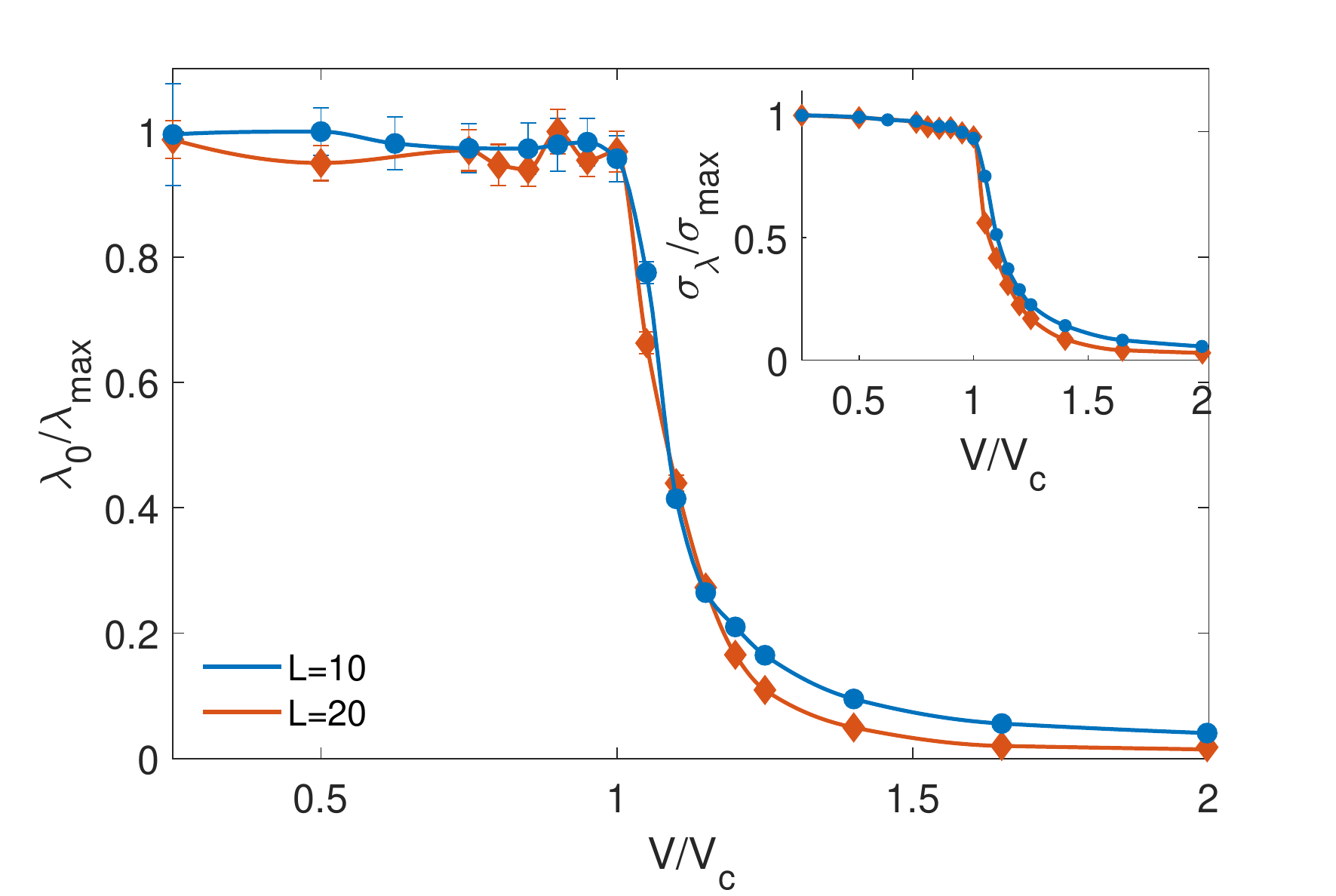}
    \caption{Normalized characteristic length $\lambda_0$ (main panel) and standard deviation $\sigma_\lambda$ (inset) across the superfluid to checkerboard phase transition ($(V/t)_c=2$) for $L=10$ (blue) and $L=20$ (orange) at inverse temperature $\beta/t=18$ and $\beta/t=36$ respectively.
     }
    \label{fig:sig_xi}
\end{figure}

\emph{Superfluid to checkerboard phase transition}
We perform simulations of the extended Bose-Hubbard model in the square lattice with $H_0=V\sum_{\langle ij\rangle}n_i n_j$ for a range of $V/t$, at fixed filling factor $1/2$, $\beta/t=18$ ($36$), and system sizes $L=10$ ($20$). In the upper row of Fig.~\ref{fig:pL_distr}, we plot distribution $p_\lambda$ for various values of $V/t$. The SF-CB transition occurs at $(V/t)_c=2$~\cite{Hebert2001}. We observe that, upon entering the CB phase, the tail of the distribution decays exponentially to zero, while in the SF phase, the distribution remains finite approaching $\lambda=N\beta$. By fitting the exponential tail (orange lines) we extract the characteristic length. In Fig.~\ref{fig:sig_xi} we plot normalized $\lambda_0$ (main plot) and $\sigma_\lambda$ (inset) as a function of $V/t$. Both $\sigma_\lambda$ and $\lambda_0$ are finite in the SF phase and decay to zero in the CB phase. The transition point is clearly marked by a significant drop in both quantities.

\begin{figure}
    \centering
    \includegraphics[width=1\columnwidth]{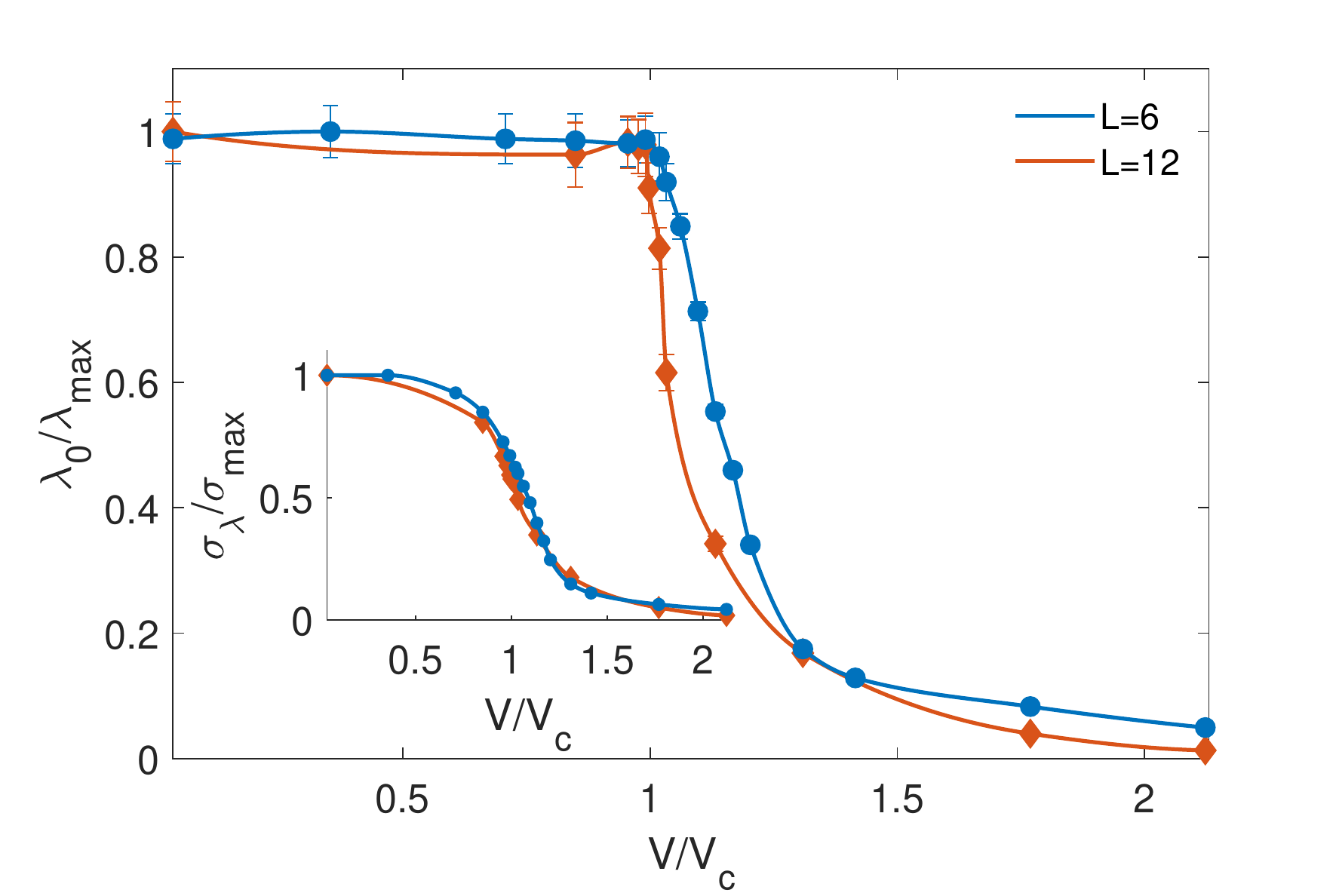}
    \caption{Normalized characteristic length $\lambda_0$ (main panel) and standard deviation $\sigma_\lambda$ (inset) across the superfluid to $\mathbb{Z}_2$ topologically ordered insulator transition ($(V/t)_c=7.0665$) for $L=6$ unit cells (blue) and $L=12$ unit cells (orange) at inverse temperature $\beta/t=18$ and $\beta/t=36$ respectively. }
    \label{fig:sig_xiZ2}
\end{figure}

\emph{Superfluid to $\mathbb{Z}_2$ topologically ordered insulator phase transition}
We perform simulations of the extended Bose-Hubbard model with $H_0=V\sum_{\hexagon}n_i n_j$, where the sum over $\hexagon$ refers to the sum between sites on the same hexagon of the Kagome lattice. We study the model for a range of $V/t$, at fixed filling factor $1/2$, $\beta/t=18$ ($36$), and system sizes $L=6$ ($12$) unit cells. In the lower row of Fig.~\ref{fig:pL_distr}, we plot distribution $p_\lambda$ for various values of $V/t$. The SF to $\mathbb{Z}_2$ topologically ordered insulator transition occurs at $(V/t)_c=7.0665$~\cite{Isakov2011}. We observe that, upon entering the $\mathbb{Z}_2$ topologically ordered insulator phase the tail of the distribution decays exponentially to zero, and, for larger interaction, nontrivial features in the form of a secondary peak at $\lambda=5$ appear in the distribution. In the SF phase, on the other hand, the distribution remains finite approaching $\lambda=N\beta$. Orange lines are fits to exponential tails from which we extract the characteristic length $\lambda_0$. Likewise in the CB-SF transition, one can infer the transition point by plotting $\lambda_0$ (main plot of Fig.~\ref{fig:sig_xiZ2}) or $\sigma_\lambda$ (inset of Fig.~\ref{fig:sig_xiZ2}) as a function of $V/t$. These plots are very similar to the ones in Fig.~\ref{fig:sig_xi} and the transition point is clearly marked by a significant drop in both quantities.

\section{Conclusions}\label{concl}
In this paper, we have studied the statistics of permutation cycles for quantum phases stabilized by several Bose-Hubbard-type models on both square and Kagome lattice. We have shown that it is generally possible to differentiate ground-states in terms of the probability distribution of the length of permutation cycles. Moreover, we have observed a secondary peak in distributions of insulating phases stabilized on the Kagome lattice. These peaks are consistent with local resonances of occupation numbers in the ground-state expansion, and can be explained heuristically as a result of the interplay between interaction, hopping and filling-factor within a first-order perturbative approach. We have also confirmed that suitable quantities characterizing distribution $p_\lambda$ can be used to find the superfluid to insulating phase transition.

\section{Acknowledgements}
All the authors wish to thank Lorenza Viola and Vittorio Penna for fruitful discussions about the manuscript.
\\\\
\appendix
\section{Weights of Path-Integral configurations}\label{appA}
According to the path-integral representation of the density matrix \cite{Ceperley1984,Ceperley1995,Prokofev1998,multiworm}, a configuration $\chi$ is defined as a specified sequence of imaginary-time instants $0<\tau_1<\ldots<\tau_{n-1}<\beta$ and corresponding Fock states $\{\ket{\alpha},\ket{\alpha_1},\ldots,\ket{\alpha_{n-1}},\ket{\alpha}\}$. Within this picture, the weight $\omega_\chi$ of equation (\ref{eq3}) are expressed as
\begin{equation}
\omega_\chi\propto(-1)^n\mel{\alpha}{H_1(\tau_1)}{\alpha_1}\cdots\mel{\alpha_n}{H_1(\tau_n)}{\alpha} \; .
\label{weights}
\end{equation}
where $H_1(\tau_\ell)= -t\sum_{i_{\ell}\,j_{\ell}} e^{E_{\ell}\tau_{\ell}}a^\dag_{i_{\ell}}a_{j_{\ell}}e^{-E_{\ell+1}\tau_{\ell}}$ and, more specifically
\begin{equation}
    \omega_\chi\propto (-1)^{2n}\,t^n \prod_\ell^{n-1} \sqrt{n_{i_\ell}+1}\sqrt{n_{j_{\ell}}} \,e^{-E_{\ell +1}d\tau_{\ell\,\ell+1}}
\end{equation}
where $E_{\ell+1}$ is the diagonal energy of state $\ket{\alpha_{\ell+1}}$, and $d\tau_{\ell\,\ell+1}$ is imaginary-time distance between two consecutive states $\ket{\alpha_\ell}$ and $\ket{\alpha_{\ell+1}}$ in the configuration. Note that for hardcore bosons the terms $\sqrt{n_{i_\ell}+1}\sqrt{n_{j_{\ell}}}$ reduce to $1$.

\section{Finite-size effects of $p_\lambda$ in the superfluid phase}\label{appB}

\begin{figure}
    \centering
    \includegraphics[width=1\columnwidth]{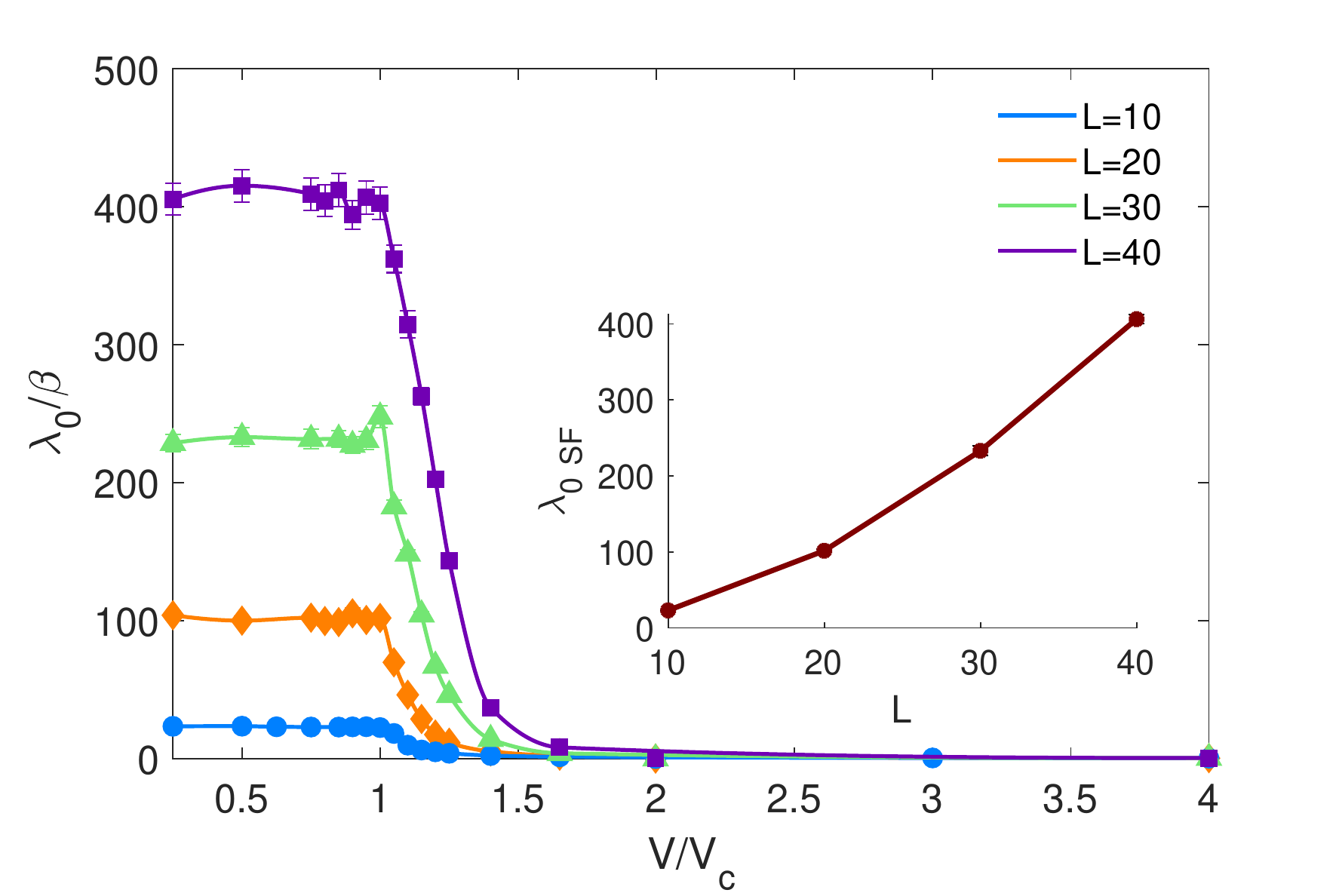}
    \caption{Characteristic length $\lambda_0$ across the superfluid to checkerboard phase transition ($(V/t)_c=2$) for $L=10$ (blue), $L=20$ (orange), $L=30$ (green) and $L=40$ (violet).
    Inset shows the increment of $\lambda_0$ within the SF phase ($V/t<2$) as a function of system size $L$. $\lambda_{0\;SF}$ is computed by averaging the values of $\lambda_0$ within the superfluid phase (for $V/t<(V/t)_c$).
     }
    \label{fig:xi_L}
\end{figure}

\begin{figure}
    \centering
    \includegraphics[width=1\columnwidth]{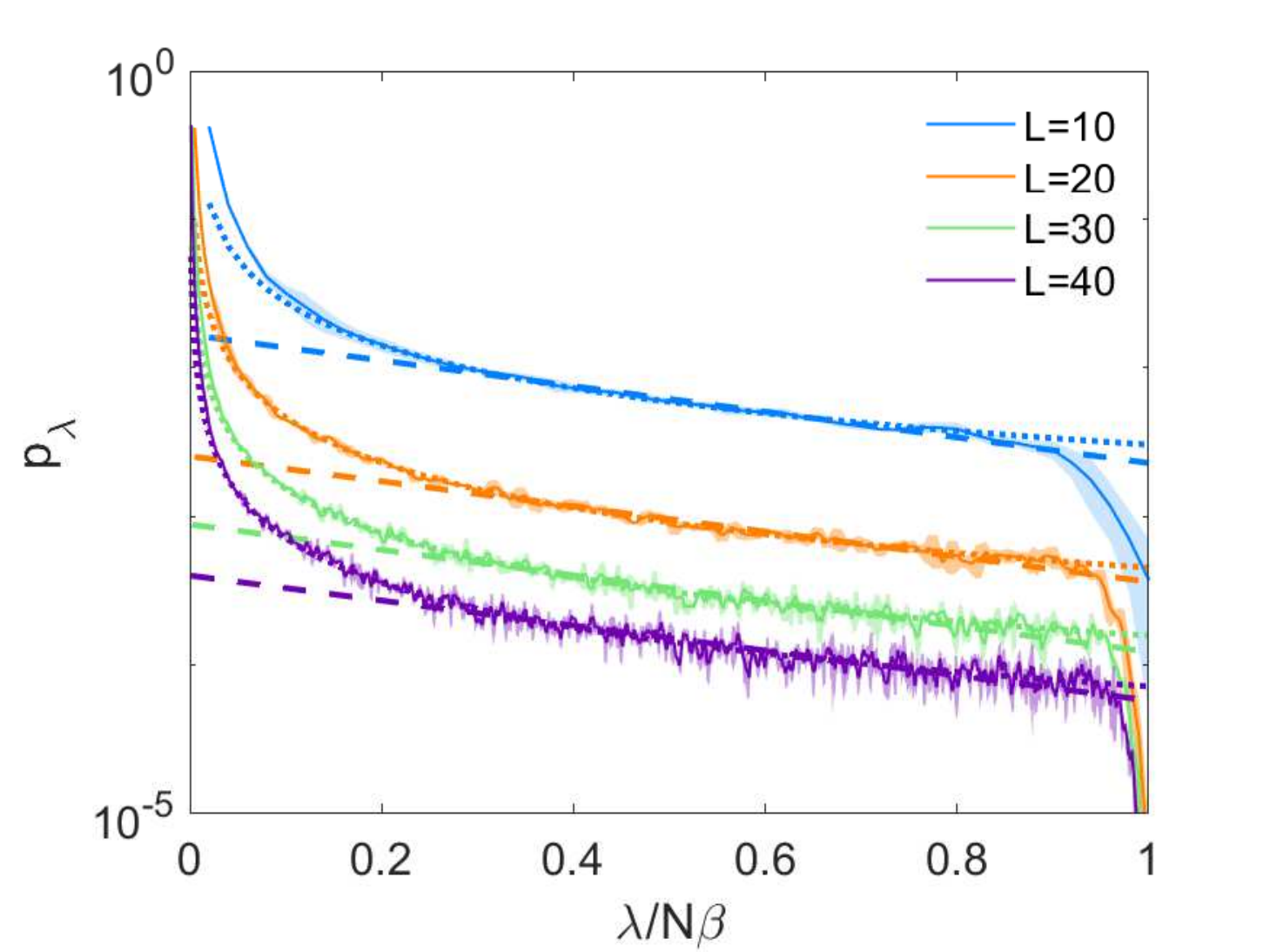}
    \caption{Permutation cycle distribution $p_\lambda$ at $V/t=1$, half-filling and  $\beta=3L/2$, for different system sizes: $L=10$ (blue), $L=20$ (orange), $L=30$ (green) and $L=40$ (violet).
    Dotted lines show the polynomial fits $b_1\cdot\lambda^{b_2}$ of $p_\lambda$ ($b_2\simeq-1$). Dashed lines represent the exponential fits $\lambda_1\cdot e^{-\lambda/\lambda_0}$ of the tail of $p_\lambda$.
     }
    \label{fig:pL_L}
\end{figure}

We note that in the entire superfluid region the characteristic length $\lambda_0$ for a given system size $L$ stays constant as a function of $V/t$. We also notice that, in the superfluid phase, distribution $p_\lambda$ becomes flatter  as the system-size increases.
This is consistent with an increasing value of $\lambda_0$ as a function of L, as  shown in Fig.~\ref{fig:xi_L}. In the inset of Fig.~\ref{fig:xi_L}, we show the divergence of $\lambda_0$ with $L$. Another possible behavior describing  $p_\lambda$ is a power law decay. In Fig.~\ref{fig:pL_L}, we plot distribution $p_\lambda$ along with the power law  fit $b_1\cdot x^{b_2}$ (dashed line)  and the exponential fit $\lambda_1\cdot e^{-\lambda/\lambda_0}$ (dotted line). As it can be noticed, the power-law fit, featuring a characteristic exponent $b_2\simeq-1$, is capable of describing the $p_\lambda$ distribution in the SF phase within our errorbars. Nonetheless, because within our errorbar, the tail of the distribution can also be described by an exponential decay, in analogy with what done in the insulating phase, we use the exponential fit to extract the characteristic length $\lambda_0$. Such characteristic length  can be used to pinpoint the phase transition.

\end{CJK*} 

\bibliography{Refs.bib}
\end{document}